# Is the degree of saturation a good candidate for Bishop's $\chi$ parameter?

J.M. Pereira & O. Coussy
*Université Paris-Est, UR Navier, École des Ponts ParisTech, Marne-la-Vallée, France.*

E.E. Alonso, J. Vaunat & S. Olivella
*Universitat Politècnica de Catalunya, Department of Geotechnical Engineering and Geosciences, Barcelona, Spain.*

ABSTRACT: In unsaturated soil mechanics, the quest for an effective stress playing the same role as Terzaghi's effective stress does for saturated soils has introduced a long standing debate, dating back to the 1960s. Several contributions have been proposed since the early work of Bishop. It is well recognized to date that a single constitutive stress is not sufficient by itself to catch the main features of the behaviour of unsaturated soils and it is often combined with matric suction. In this paper, focus is given to a largely used formulation for such a constitutive stress, based on the use of an averaged pore pressure. In particular, this paper discusses on thermodynamics bases the validity of the choice of the factor $\chi$ weighting the fluid pressures contribution to the constitutive stress. This factor is usually assumed to be equal to the degree of saturation of water. In this work it is shown that the choice of this natural candidate implies restrictive assumptions on the plastic flow rule. As shown from experimental data obtained from a literature review, this choice may not be pertinent for certain classes of materials, particularly high plasticity clays.

## 1 INTRODUCTION

In engineering applications involving mechanics of geomaterials, focus is generally put on the deformation of the solid skeleton so that a great interest is given to the definition of the stresses governing the skeleton deformation. The presence of one or several fluids within the porous space of these materials introduces a complexity in the choice of the stress state variables with respect to non-porous materials. (Terzaghi, 1936) stated that in the case of saturated soils (where a single fluid filling the porous space) a single stress (called effective stress) governs the deformation and strength.

Roscoe and his co-workers (Roscoe et al., 1958) applied the mathematical theory of plasticity to saturated soil mechanics introducing the concept of critical state. This work resulted in the well-known Cam-clay model for saturated soils involving Terzaghi's effective stress. An extension of this model to unsaturated soils has been proposed by Alonso et al. (Alonso et al., 1990) within a simple elastoplastic framework. This extension points out the need of two independent state variables to capture the experimentally observed behaviour of unsaturated soils instead of a unique stress variable (effective stress) in the case of saturated states.

Later on, this model settled the bases of numerous models, addressing additional features of unsaturated soils behaviour, such as the effects of water content (Wheeler, 1996) or degree of saturation (Jommi and Di Prisco, 1994, Dangla et al., 1997, Lewis and Schrefler, 1998, Gallipoli et al., 2003, Sheng et al., 2004, Pereira et al., 2005) among others. Advances on the last point have reintroduced a strong debate dated back to the 1960s with Bishop's proposal for an extended effective stress **σ**' to unsaturated states (Bishop, 1959) who introduced an averaging parameter $\chi$, this latter being a function of the water degree of saturation $S_r$:

$$\boldsymbol{\sigma}' = \boldsymbol{\sigma} - u_a \mathbf{1} + \chi(S_r) s \mathbf{1} \qquad (1)$$

where **σ** is the total stress tensor, $u_a$ is the air pressure and $s = u_a - u_w$ the suction. In recent years, the most commonly used assumption for this parameter $\chi$ is to assume $\chi = S_r$.

In the first part of this paper, the thermodynamics of plasticity for unsaturated geomaterials is revisited with a particular attention paid to the constitutive stress governing the skeleton deformation. It is shown that Bishop's proposal with the common assumption $\chi = S_r$ rely on a restrictive assumption concerning the plastic flow rule. In a second part of the paper, experimental evidences of an existing deviation to the $\chi = S_r$ assumption are presented. These evidences are interpreted on the basis of microstructural considerations.

## 2 THERMODYNAMICS OF PLASTICITY

When considering an unsaturated soil, the analysis of the contribution of each fluid phase to the strain work of the solid skeleton (composed of the solid matrix together with the interfaces) is not as straightforward as it may be when the soil is saturated. In this case, changes of the Lagrangian partial porosities filled by each fluid phase $d\phi_a$ and $d\phi_w$ are due not only to changes of the porous volume occupied by each phase but also to the invasion of the volume previously containing one phase by the other phase. To overcome this difficulty, the partial porosities are split into two parts, thus separating invasion and deformational processes:

$$\phi_w = s_r \phi = S_r \phi_0 + \varphi_w \qquad (2)$$

$$\phi_a = (1 - s_r) \phi = (1 - S_r) \phi_0 + \varphi_a \qquad (3)$$

$$\phi = \phi_0 + \varphi_a + \varphi_w \qquad (4)$$

where $s_r$, $S_r$, $\phi$, $\phi_0$, $\varphi_w$ and $\varphi_a$ respectively stand for the Eulerian and Lagrangian degrees of saturation, (Coussy, 2007, Coussy et al., 2009), the porosities in the actual configuration and in the reference configuration and the changes due to deformation only of the Lagrangian porosity for the part of the porous network occupied by water and air. It should be noted that this partition can not be obtained when Eulerian variables are used. In this latter case, changes in $s_r$ may be indifferently due to both drying/wetting and deformation processes. Use of Eqs. (2)–(4) allows expressing the work input as:

$$dw = d\omega + d\theta \qquad (5)$$

$$d\omega = p d\varepsilon_v + u_a d\varphi_a + u_w d\varphi_w + q d\varepsilon \qquad (6)$$

$$d\theta = -\phi_0 (u_a - u_w) dS_r \qquad (7)$$

According to Eqs. (5)–(7), the work input is split into the contribution needed to deform the skeleton, $d\omega$, and that required for the invasion process to occur, $d\theta$. By noting $U$ the fluid-solid interface energy per unit of initial volume, Eq. (7) allows for stating:

$$dU = -\phi_0 (u_a - u_w) dS_r \qquad (8)$$

This relation implies that $U$ and also suction $u_a - u_w$ must be functions of $S_r$ only so that $u_a - u_w = r(S_r)$, which is the classical expression of the water retention curve. This relation does not account for hysteretic effects. Accounting for them is not contradictory to the approach presented here (see e.g. (Dangla et al., 1997)), but requires the consideration of appropriate energy couplings that would weight down the text.

Since the considered system is a closed system (fluids have been removed), Clausius-Duhem inequality which contains the first and the second laws of thermodynamics expresses that, for isothermal evolutions, the strain work input $dw$ to the system has to be greater or equal to the infinitesimal free energy $dF$ that the system can store, the difference $dD$ being spontaneously dissipated into heat. Assuming the solid grains incompressibility and using Eqs. (5)–(7), Clausius-Duhem inequality writes as follows:

$$dD = (p - u_a) d\varphi_a + (p - u_w) d\varphi_w + q d\varepsilon_q \\ - \phi_0 (u_a - u_w) dS_r - dF \geq 0 \qquad (9)$$

Any further analysis requires the assumptions on the dependency of the free energy $F$ on the state variables. Denoting $\Psi$ the elastic energy and $Z$ the locked energy (Collins, 2005), the following dependencies are assumed:

$$F = \Psi(\varphi_a - \varphi_a^p, \varphi_w - \varphi_w^p, \varepsilon_q - \varepsilon_q^p, S_r) + Z(S_r, \alpha) \\ + U(S_r) \qquad (10)$$

The state equations are obtained considering elastic evolutions thus leading to null plastic strains and an equality in (9). They read:

$$p - u_a = \frac{\partial \Psi}{\partial \varphi_a}; \quad p - u_w = \frac{\partial \Psi}{\partial \varphi_w}; \quad q = \frac{\partial \Psi}{\partial \varepsilon_q}; \\ \phi_0 (u_a - u_w) = -\frac{\partial(\Psi + Z)}{\partial S_r} - \frac{dU}{dS_r} \qquad (11)$$

The first three sub-equations capture to the elastic behaviour of the solid matrix. The last one corresponds to the expression of the water retention curve. It includes the effects of the deformation of the porous volume, except those leading to hysteretic phenomena as stated before. Using state equations (11) and assumptions on dependencies of $F$ given by equation (10), Clausius-Duhem inequality is expressed as:

$$dD = (p - u_a) d\varphi_a^p + (p - u_w) d\varphi_w^p + q d\varepsilon_q^p \\ + \beta d\alpha \geq 0 \qquad (12)$$

$$\beta = -\frac{\partial Z(S_r, \alpha)}{\partial \alpha} \qquad (13)$$

$\beta$ is the energy conjugate of the hardening variable $\alpha$, depending on $S_r$. It will be called hardening force and subsequently associated to the limit of elasticity. Introducing the plastic incompressibility of the solid grains which expresses as:

$$d\varphi_a^p = -(1 - \chi) d\varepsilon_v^p \; ; \; d\varphi_w^p = -\chi d\varepsilon_v^p \qquad (14)$$

it comes from (12):

$$dD = p^B d\varepsilon_v^p + q d\varepsilon_q^p + \beta d\alpha \qquad (15)$$

where $p^B$ is Bishop's stress defined as:

$$p^B = p - [1 - \chi(S_r)] u_a - \chi(S_r) u_w \qquad (16)$$

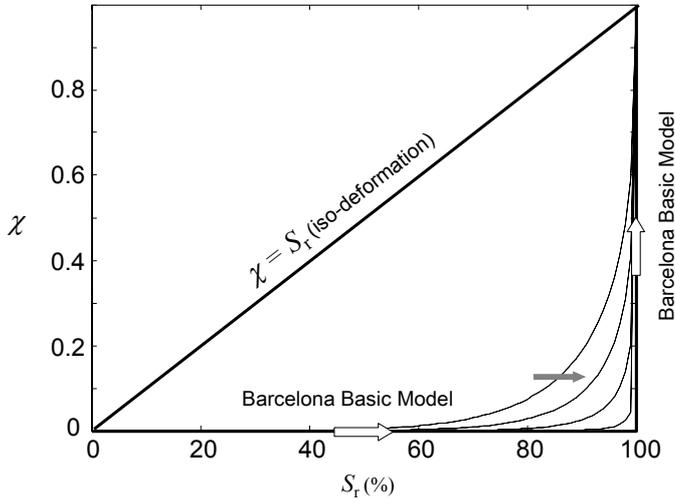

Figure 1: Family of functions χ($S_r$) suitable for entering in the definition of an extended Bishop's effective stress. Linear function χ($S_r$) = $S_r$ define the classical Bishop's stress, step function χ(Sr < 1) = 0, χ (1) = 1 the pair net stress / Terzaghi's effective stress.

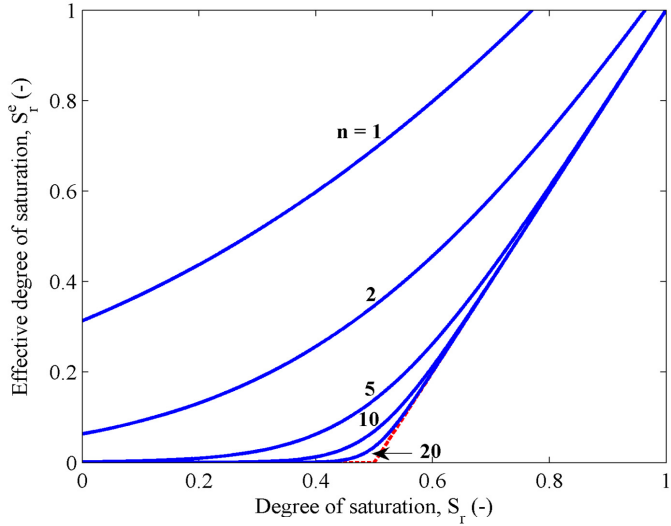

Figure 2: Effective degree of saturation as a function of the total degree of saturation based on the exponential smoothing function; influence of $n$ for $S_r^m = 0.5$.

Figure 1 shows Family of functions χ($S_r$) suitable for entering in the definition of an extended Bishop's effective stress. Interestingly, assuming isodeformation of the porous volumes occupied by water and air, that is:

$$d\varphi_w^p / \phi_0 S_r = d\varphi_a^p / \phi_0 (1 - S_r) \qquad (17)$$

and comparing (14) and (17) leads to χ($S_r$) = $S_r$, which is actually the commonly used assumption. According to (15), Bishop's stress plays the same role in unsaturated states as does Terzaghi's proposal ($p' = p - u$) in saturated conditions. Actually, both definitions arise from solid grains incompressibility assumptions.

## 3 MICROSTRUCTURAL INTERPRETATION

The previous study has lead to the conclusion that some care must be taken when using the so-called average constitutive stress (equivalent to Bishop's stress where χ factor is assumed to be equal to the degree of saturation $S_r$). The remaining part of the paper is devoted to the description of a proposal with the aim of providing a better description of the suction contribution to the constitutive stress.

Starting from common observations on pore size distributions of soils, it may be argued that two classes of pores are generally distinguishable. The first class corresponds to the largest pores (macropores). The fraction of the water filling these pores will be denoted $S_r^M$. The second class corresponds to the smallest pores (micropores). The fraction of the water filling these pores will be denoted $S_r^m$. It is obvious that the availability of the water filling one or the other class of pores will be different. For the first class, water exchanges are mainly governed by capillary effects. For the second class, water is quite more attached to the solids by physico-chemical interactions. This part of the water is not so freely available. As an illustration, some authors have reported experimental observations showing that this water does not participate to the darcean transport of water, thus leading to a reduced apparent permeability of the soil (Romero, 1999).

Based on these microstructural considerations, it is proposed to assume that χ factor is no more equal to the 'total' degree of saturation but to an 'effective' degree of saturation defined as follows:

$$S_r^e = \left\langle \frac{S_r - S_r^m}{1 - S_r^m} \right\rangle \qquad (18)$$

where $\langle x \rangle = 1/2(x + |x|)$ represents Macaulay brackets. Such an assumption on χ factor results in the following expression for the constitutive stress:

$$\boldsymbol{\sigma}' = \boldsymbol{\sigma} - p_g \mathbf{1} + S_r^e s \mathbf{1} \qquad (19)$$

Of course, the rough separation of the pore sizes presented above may not be so sharp in reality. Furthermore, it may be interesting to avoid the second order discontinuity at $S_r = S_r^m$. This is particularly true when dealing with numerical analyses. Smoothing techniques for the "corner" of the piece-wise proposal have been examined. The following expression for $S_r^e$:

$$S_r^e = \frac{S_r - S_r^m}{1 - S_r^m} + \frac{1}{n} ln \left[ 1 + exp\left( -n \frac{S_r - S_r^m}{1 - S_r^m} \right) \right] \qquad (20)$$

provides a smoothing of the corner, which is controlled by parameter $n$. As $n$ increases, Eq. (20) becomes closer to the piecewise approximation (see Figure 2). In order to avoid the introduction of a new parameter and since no direct physical meaning may be attributed to $n$, it is advised to fix the value of this smoothing parameter $n$. A high value, for instance $n = 20$ may be used in practice. (Alonso et al., 2009) proposed another smoothing function.

## 4 VALIDATION AND DISCUSSION

The validation of this proposal is checked on the basis of elastic stiffness and shear strength data. A set of soils have been chosen, covering a large range of soil types from mostly granular materials to high plasticity clays. In what follows, all the water retention curves have been modelled using experimental data available in the literature to fit the modified van Genuchten model MVG proposed by (Romero and Vaunat, 2000):

$$S_r(s) = \left(1+(\alpha_l s)^n\right)^{-m} \left[1 - \frac{\ln\left(1+\frac{s}{s_{res}}\right)}{\ln\left(1+\frac{a}{s_{res}}\right)}\right] \quad (21)$$

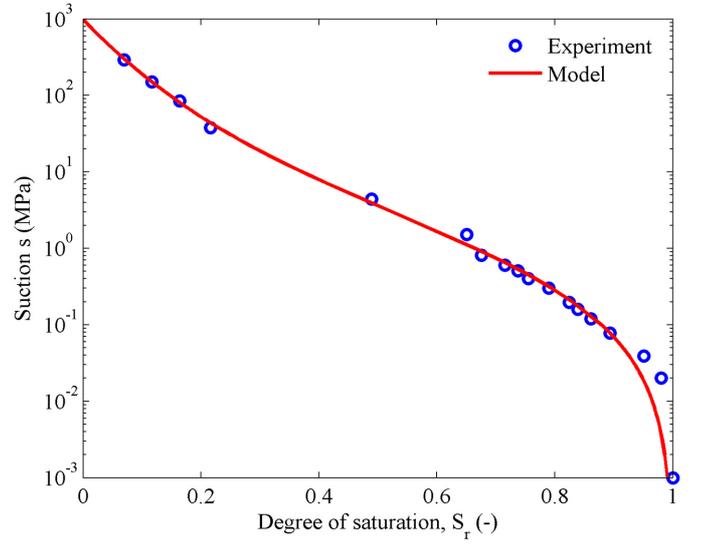

Figure 3: Water retention curve of a glacial till from Canada: experimental data after (Vanapalli et al., 1996) and simulated curve using modified van Genuchten model.

### 4.1 From shear strength data

Substitution of equation (19) into the classical Mohr-Coulomb criterion in terms of effective stress gives:

$$\tau = \left[c' + S_r^e\, s \tan\varphi'\right] + (\sigma - p_g)\tan\varphi' \quad (22)$$

where the first term between brackets in the right hand side corresponds to the apparent cohesion due to suction. Shear strength data of a sandy-silt from Switzerland (Geiser et al., 2006), a glacial till from Canada (Vanapalli et al., 1996) and a decomposed stuff from Hong Kong (Fredlund et al., 1996) were studied. Experimental data were fitted using Eq. (22) and adjusting parameter $S_r^m$.

As an illustration and for the sake of conciseness, only the case of the glacial till from Canada is presented here. Figure 3 shows the water retention curve of this soil (experimental data and best-fitted curve using MVG model). Figure 4 presents the evolution of the shear strength with suction. In addition to the values obtained using the proposal made in this study, shear strength data have been simulated using the assumption $\chi = S_r$. The values are reported in Fig. 5. It is obvious that these values overestimate the experimental data. Simulations using the proposal show that a good fit has been obtained.

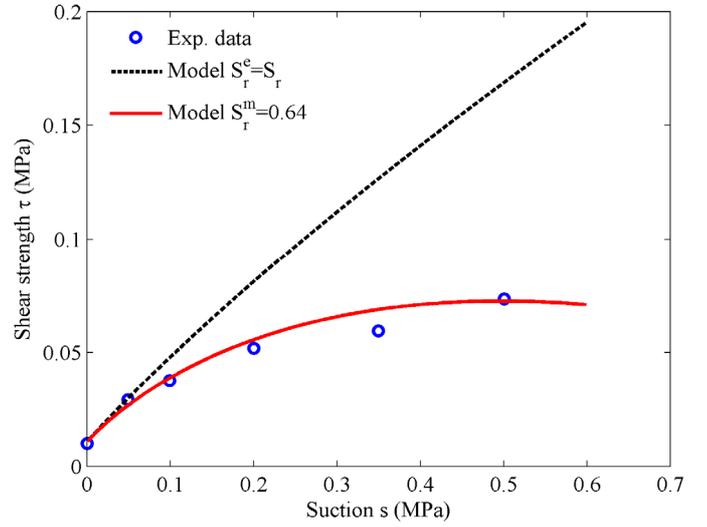

Figure 4: Shear strength of a glacial till from Canada: experimental data after (Vanapalli et al., 1996) and simulated curve using the constitutive stress proposed in this paper (Eq. 19).

### 4.2 From elastic properties

Assuming the validity of the constitutive stress for unsaturated states and using the usual elastic equation, it comes:

$$d\varepsilon_v^e = \kappa \frac{dp'}{p'} \quad (23)$$

which should be compared to the relation used in formulations using net stress as the mechanical stress variable such as in the Barcelona Basic Model (Alonso et al., 1990) and which reads as follows:

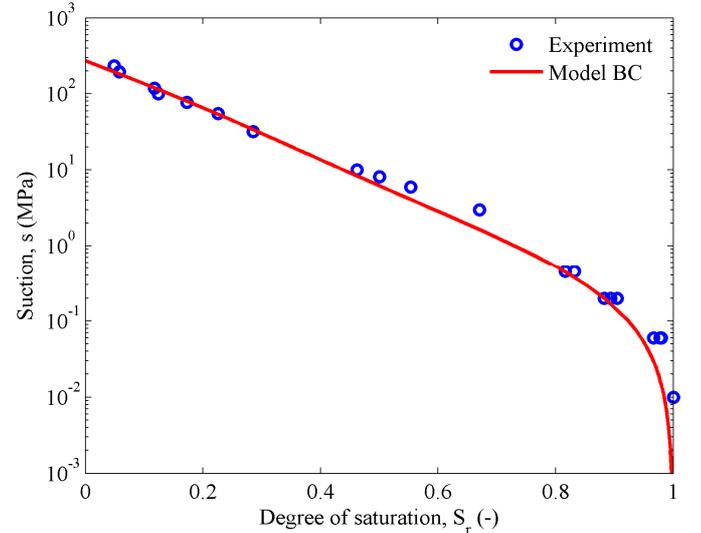

Figure 5: Water retention curve of Boom clay at a dry unit weight of 16.7 kN/m³: experimental data after (Romero, 1999) and simulated curve using modified van Genuchten model.

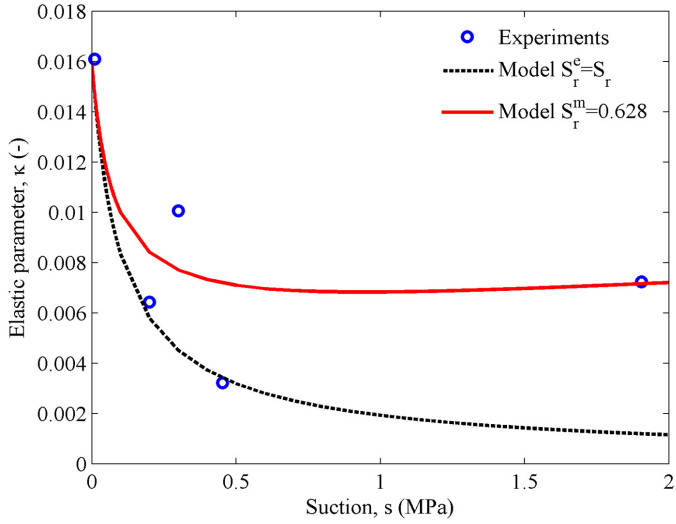

Figure 6: Elastic parameter $\kappa$ for Boom clay at a dry unit weigth of 16.7 kN/m$^3$: experimental data after (Romero, 1999) and simulated values using the constitutive stress proposed in this paper (Eq. 19).

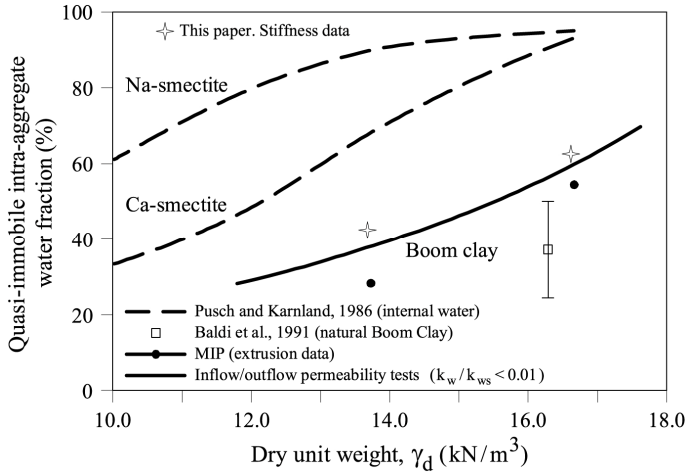

Figure 7: Quasi-immobile intra-aggregate water fraction (% of total porous volume): experimental data after (Romero, 1999).

$$d\varepsilon_v^e = \bar{\kappa}\frac{d\bar{p}}{\bar{p}} \qquad (24)$$

Various authors have reported that $\bar{\kappa}$ values were dependent on applied suction (see (Romero, 1999) for instance). It is obvious when looking at Eq. (23) that a dependence of the apparent stiffness upon suction values is directly accounted for *via* the presence of the suction in the definition of the constitutive stress. The analysis of values of the elastic parameter $\bar{\kappa}$ reported in the literature has been performed. This data has been reinterpreted in terms of effective stress using Eq. (23) with constant $\kappa$. A good fit of this data particularly of its evolution with suction may be obtained if a proper definition of the constitutive stress is considered. The data analyzed concern an aeolian deposited silt from Jossigny (France) (Cui and Delage, 1996) and a high plasticity clay from Boom (Belgium) at two different dry unit weights (respectively 13.7 and 16.7 kN/m$^3$) (Romero, 1999). A similar analysis has also been performed on the elastic shear moduli for compacted specimens of the core of Vallfornés dam. The shear moduli were obtained using resonant column tests (Alonso, 1998). As an illustration, the case of the densest Boom clay is presented. The water retention curve and the elastic parameter $\bar{\kappa}$ evolution as a function of suction for this soil are showed in Figures 5 and 6 respectively. Again, comparison with simulations using $\chi = S_r$ assumption overestimate the overall contribution of suction to the effective stress thus predicting too high values of the soil stiffness (low values of $\bar{\kappa}$).

### 4.3 *Discussion*

Table 1 summarizes the material parameters fitted from experimental water retention properties of the different soils analyzed in this paper. In all cases, the theoretical model is a variation of van Genuchten equation (MVG model). Table 2 presents the material parameter $S_r^m$ used to define the effective degree of saturation appearing in the constitutive stress instead of the usual assumption consisting in letting $\chi = S_r$. In this table, the soils are sorted in the order of increasing $S_r^m$ values. It is interesting to note that the values of the material parameter $S_r^m$ involved in the proposed definition of the effective stress are well correlated with the microstructure of the different soils accounted for in this study. Granulometric properties of these soils are summarized in Table 2. They show that increasing values of $S_r^m$ also correspond to increasing values of the content of finer solid particles. This correlation corroborates the microstructural interpretation given earlier. Indeed, three soil classes may be tentatively identified from the fitted parameters. The first class corresponds to fairly granular soils for which the amount of microscopically trapped water is negligible thus leading to the validity of the assumtion $\chi = S_r$. The second group gathers silty soils for which the amount of microstructural water represents intermediate values. The finest soils correspond to the third group and thus to the highest values of $S_r^m$. An influence of the dry unit weigth is also observed: denser materials seem to be characterized by higher values of $S_r^m$ which is coherent with our interpretation (denser material correspond to lower volumes of macropores and thus to higher fractions of micropore relatively to the total volume of pores). Another important point is illustrated in Figure 7. The amount of "quasi-immobile water" as introduced by (Romero, 1999) is plotted as a function of the dry unit weight of Boom clay. This notion is equivalent to the microstructural water $S_r^m$ used in this study. The values reported by Romero correspond to estimations obtained from completely independent techniques (mercury intrusion porosimetry (MIP) or permeability tests). The $S_r^m$ values obtained here from elastic stiffness have been plotted and nicely fit in the original plot.

# 5 CONCLUSIONS

Thermodynamics of unsaturated soils plasticity has been revisited and a thermodynamically consistent framework for hardening plasticity has been presented. It has also been shown that the choice $\chi = S_r$ rely on a restrictive assumption on the plastic flow rule. The proposed framework uses a generalized effective stress which includes both the commonly used effective stress with $\chi = S_r$ and the net stress as limiting cases. In between these limiting cases, the general evolution of $\chi$ factor with the degree of saturation may not be identified without supplementary peace of information. To this end, a microstructural interpretation of the repartition of the water phase into the porous space has been used. This proposal has been validated on the basis of experimental data from a given set of soils ranging from low to high plasticity soils. It has been concluded that the commonly used choice for $\chi$ factor that is $\chi = S_r$, must be taken with care since important deviation from this assumption appear for relatively high plasticity soils.

Table 1: Parameters of the water retention curve model for the different soils.

| Soil | $n$ | $m$ | $\alpha_I$ | $s_r$ | $A$ |
|---|---|---|---|---|---|
| | – | – | MPa$^{-1}$ | MPa | MPa |
| Decomposed tuff | 3.63 | 0.14 | 36.14 | 1000 | 1000 |
| Vallfornés dam core | 1.11 | 0.67 | 0.38 | 1000 | 1000 |
| Sion silt | 3.25 | 0.24 | 19.08 | 1000 | 1000 |
| Jossigny silt | 4.56 | 0.026 | 35.54 | 1000 | 1000 |
| Glacial till | 0.59 | 0.67 | 0.72 | 1000 | 1000 |
| Boom clay ($\gamma_d$=13.7 kN/m$^3$) | 1.14 | 0.196 | 21.29 | 274 | 274 |
| Boom clay ($\gamma_d$=16.7 kN/m$^3$) | 0.75 | 0.354 | 1.55 | 274 | 274 |

Table 2: Granulometric properties and microscopic degree of saturation of the different soils.

| Soil | Sand / Silt / Clay fractions | $S_r^m$ |
|---|---|---|
| | % | – |
| Decomposed tuff | 60 / 35 / 5 | 0.02 |
| Vallfornés dam core | 54 / 40 / 6 | 0.25 |
| Sion silt | 20 / 72 / 8 | 0.40 |
| Jossigny silt | 4 / 62 / 34 | 0.56 |
| Glacial till | 28 / 42 / 30 | 0.64 |
| Boom clay ($\gamma_d$=13.7 kN/m$^3$) | 18 / 30 / 52 | 0.42 |
| Boom clay ($\gamma_d$=16.7 kN/m$^3$) | 18 / 30 / 52 | 0.63 |